\documentclass[aps,prl,twocolumn,superscriptaddress]{revtex4-2}
\usepackage[english]{babel}
\usepackage{bm}
\usepackage{amsmath}
\usepackage[dvips]{graphicx}
\usepackage[colorlinks=true, allcolors=blue]{hyperref}
\usepackage{diagbox}
\usepackage{multirow}

\begin{document}

\title{Twirling and spontaneous symmetry breaking of domain wall networks in lattice-reconstructed heterostructures of 2D materials}

\author{M.~A.~Kaliteevsky}
\email{mikhail.kaliteevski@manchester.ac.uk}
\affiliation{National Graphene Institute, University of Manchester, Booth St. E. Manchester M13 9PL, United Kingdom}
\affiliation{University of Manchester, School of Physics and Astronomy, Oxford Road, Manchester M13 9PL, United Kingdom}

\author{V.~V.~Enaldiev}
\affiliation{National Graphene Institute, University of Manchester, Booth St. E. Manchester M13 9PL, United Kingdom}
\affiliation{University of Manchester, School of Physics and Astronomy, Oxford Road, Manchester M13 9PL, United Kingdom}

\author{V.~I.~Fal'ko}
\email{vladimir.falko@manchester.ac.uk}
\affiliation{National Graphene Institute, University of Manchester, Booth St. E. Manchester M13 9PL, United Kingdom}
\affiliation{University of Manchester, School of Physics and Astronomy, Oxford Road, Manchester M13 9PL, United Kingdom}
\affiliation{Henry Royce Institute for Advanced Materials, University of Manchester, Oxford Road, Manchester, M13 9PL, UK}

\begin{abstract}
Lattice relaxation in twistronic bilayers with close lattice parameters and almost perfect crystallographic alignment of the layers results in the transformation of moir\'e pattern into a sequence of preferential stacking domains and domain wall networks. Here, we show that reconstructed moir\'e superlattices of the perfectly aligned heterobilayers of same-chalcogen transition metal dichalcogenides have broken-symmetry structures featuring twisted nodes ('twirls') of domain wall networks. Analysing twist-angle-dependences of strain characteristics for the broken-symmetry structures we show that the formation of twirl reduces amount of hydrostatic strain around the nodes, potentially, reducing their infuence on the band edge energies of electrons and holes.
\end{abstract}
\maketitle

This study addresses a detailed analysis of domain wall networks (DWN) which form in long-period moir\'e patterns characteristic for highly aligned heterostructures of same-chalcogen transition metal dichalcogenides MoX$_2$/WX$_2$ (TMDs with X=S or Se). As same chalcogen TMDs have very close lattice constants, moir\'e patterns at their interface have long periods offering a sufficient space for creating preferential stacking areas (domains). That is, the energy gain due to better adhesion can surmount the cost of intralayer strain in each of the constituent crystals. The reconstruction of small-angle twisted bilayers into an array of domains \cite{Enaldiev_PRL} has been observed \cite{Weston2020,rosenberger2020,Sung2020,McGilly2020,Shabani2021} both in MoS$_2$/WS$_2$ and MoSe$_2$/WSe$_2$ heterostructures. The observed \cite{Weston2020,rosenberger2020,Sung2020,McGilly2020,Shabani2021} and theoretically modelled \cite{NaikPRL2018,CarrPRB2018,Enaldiev_PRL} structures feature hexagonal for anti-parallel (AP) orientation of unit cell and triangular for parallel (P) orientation of unit cells  DWN.

\begin{figure}
\includegraphics[width=1.0\columnwidth]{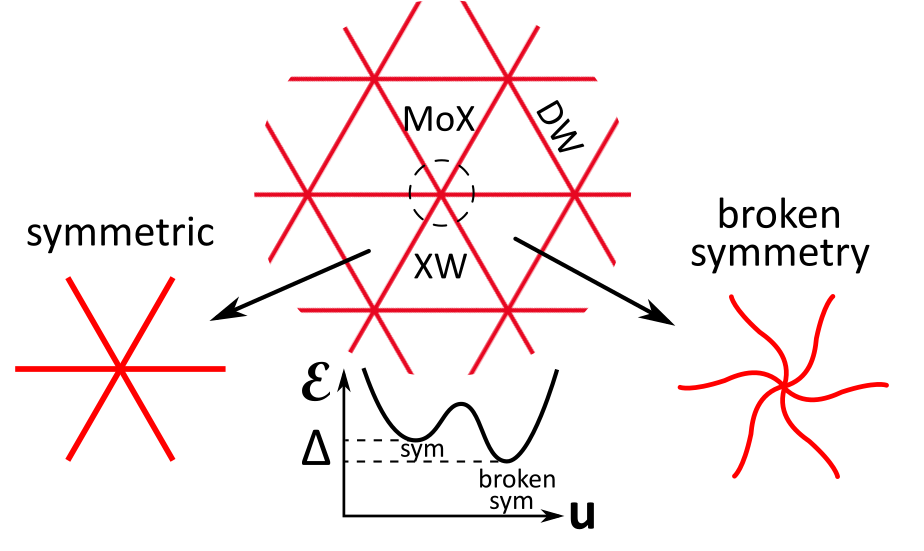}
\caption{\label{fig:1} 
Symmetric and broken-symmetry (twirl) structures of DWN in aligned P-MoX$_2$/WX$_2$ bilayers. Bottom plot shows energy gain ($\Delta$) from the formation of twirls. } 
\end{figure}

Here, we show that lattice relaxation in P/AP-MoX$_2$/WX$_2$ (X=S,Se) bilayers in the limit of $\theta=0^{\circ}$ twist angles undergoes symmetry breaking of the domain wall patterns resulting in formation of twirled structures of DWN nodes (see Fig. \ref{fig:1} for P-MoX$_2$/WX$_2$ bilayers). The emergence of twirled DWN nodes is specific for heterobilayers, in contrast to previously studied marginally twisted homobilayers (MX$_2$/MX$_2$) \cite{NaikPRL2018,CarrPRB2018,Enaldiev_PRL}, where only symmetric star-like nodes have been found. This difference stems from the hydrostatic strain component determined by a small lattice mismatch of the two constituent 2D crystals. By adjusting lattice constants of MoX$_2$ and WX$_2$ inside the domains, each monolayer compression/expansion is inflicted onto domain  walls creating hot spots of hydrostatic strain in the DWN nodes with XX (chalcogen over chalcogen) stacking \cite{Enaldiev2022a}. An excessive energy costs of a large hydrostatic strain around XX nodes (as compared to shear deformations dominating DWN structure in homobilayers) is, then, negotiated by left/right-handed twirling of the domain walls, leading to a broken-symmetry configuration sketched in Fig. \ref{fig:1}.

{\it Methods.}
To study lattice relaxation in MoX$_2$/WX$_2$ bilayers we use multiscale modelling approach developed in Refs. \cite{Enaldiev_PRL,Enaldiev_2021}. This combines mesoscale elasticity with analytically interpolated description of adhesion energy of two layers, computed using density functional theory (DFT). The latter is represented below upon taken into account interlayer distance relaxation (see SM):
\begin{equation}\label{Eq:adhesion}
\begin{split}
     W_{\rm P/AP}(\bm{r}_0) = \qquad\qquad\qquad\qquad\qquad\qquad\qquad\qquad\qquad\\
    \sum_{n,l=1,2,3}\left[w^{(s)}_{n}\cos\left(\bm{G}^{(n)}_l\bm{r}_0\right) + w^{(a)}_{n}\sin\left(\bm{G}^{(n)}_l\bm{r}_0+\gamma_{\rm P/AP}\right)\right].
\end{split}
\end{equation}
Here, $\bm{r}_0$ is a locally defined in-plane vector determining stacking arrangement between layers ($\bm{r}_0=\bm{0}$ for XX stacking, $\bm{r}_0=(0,-a/\sqrt{3})$ for XW/2H stacking and $\bm{r}_0=(0,a/\sqrt{3})$ for MoX/MoW stacking in P/AP-heterostructures), phases $\gamma_{\rm P}=\pi/2$, $\gamma_{\rm AP}=0$ account $\bm{r}_0\to-\bm{r}_0$ symmetry of the adhesion energy in P- and AP-bilayers \cite{Enaldiev_PRL}; $G_{1,2,3}^{(1,2,3)}$ are three sets of the shortest reciprocal lattice vectors ($|G_{1,2,3}^{(1)}|=G$, $|G_{1,2,3}^{(2)}|=G\sqrt{3}$, $|G_{1,2,3}^{(3)}|=2G$) of a commensurate MoX$_2$/WX$_2$ heterostructure, related by 120$^\circ$-rotations within each set. The values of parameters $w^{(s,a)}_{1,2,3}$ established using DFT \cite{Enaldiev_PRL,Enaldiev_2021} are listed in Table \ref{Tab:data}. 

Next, we combine \eqref{Eq:adhesion} with elasticity theory by substituting local stacking in the form,
\begin{equation}\label{Eq:r0}
\bm{r}_0(\bm{r})=\delta\cdot\bm{r}+\theta\hat{z}\times\bm{r} + \bm{u}^{\rm Mo}-\bm{u}^{\rm W}, 
\end{equation}
where $\delta\approx0.2\%$ for MoS$_2$/WS$_2$ bilayers and $\delta\approx0.4\%$ for MoSe$_2$/WSe$_2$; $\bm{u}^{\rm Mo}(\bm{r})$ and $\bm{u}^{\rm W}(\bm{r})$ are in-plane displacement fields describing lattice relaxation in MoX$_2$ and WX$_2$ layers, respectively. Then, we minimise total energy of moir\'e superlattice,
\begin{widetext}
\begin{equation}\label{Eq:min_functional}
\mathcal{E}=\int d^{2} \boldsymbol{r}\left\{W_{\rm AP/P}\left(\bm{r}_0(\bm{r})\right) + 
 \sum_{l={\rm W,Mo}}\left[\frac{\lambda_l+\mu_l}{2}\left({\rm div}\bm{u}^{(l)}\right)^{2} + \frac{\mu_l}{2}\left(\left(u_{xx}^{(l)}-u_{yy}^{(l)}\right)^2+4u_{xy}^{(l)2}\right)\right]\right\},
\end{equation}
\end{widetext}
with respect to the displacement fields $\bm{u}^{\rm W}$ and $\bm{u}^{\rm Mo}$. Here,  $\lambda_{\rm Mo,W}$, $\mu_{\rm Mo,W}$ are elastic moduli of MoX$_2$ and WX$_2$ monolayers (see Table \ref{Tab:data}), and \mbox{$u_{ij}^{l}=(\partial_ju^{l}_{i}+\partial_iu^{l}_{j})/2$} are components of strain tensors. 

{\setlength{\tabcolsep}{2.5pt}
\begin{table}[!h]
\begin{center}
{\small	\caption{Adhesion \cite{Enaldiev_PRL} and elastic \cite{iguiniz2019,androulidakis2018} energy parameters (in eV/nm$^2$) used in Eq. \eqref{Eq:min_functional}. \label{Tab:data}}}
	{\small
	\begin{tabular}{c|cccccc}
		\hline
		\hline
        X& \mbox{$w_{1}^{(s)}$} & \mbox{$w_{1}^{(a)}$} & \mbox{$w_{2}^{(s)}$} & \mbox{$w_{2}^{(a)}$} & \mbox{$w_{3}^{(s)}$} & \mbox{$w_{3}^{(a)}$} \\
        \hline
        S, AP & 0.1415 & 0.0269 & -0.0338 & 0 & -0.0166 & -0.0030\\
        S, P &  0.1559 &  0 & -0.0398  &  0  & -0.0199 & 0\\
        \hline
        Se, AP &  0.1128 &  0.0256 & -0.0201  &  0  & -0.0098 & -0.0024\\
        Se, P &  0.1284 &  0 & -0.0248  &  0  & -0.0124 & 0\\
        \hline
        \hline
       X & $\lambda_{\rm W}$ & $\lambda_{\rm Mo}$ & $\mu_{\rm W}$ & $\mu_{\rm Mo}$ &  & \\
        \hline
        S & 328.2  & 520.1 & 453.2 & 443.0 &  &\\
        Se & 185.5  & 264.3 & 302.6 & 310.3 & & \\
		\hline
		\hline
	\end{tabular}
	}
	\end{center}
\end{table}
}

\begin{figure*}
\includegraphics[width=2.0\columnwidth]{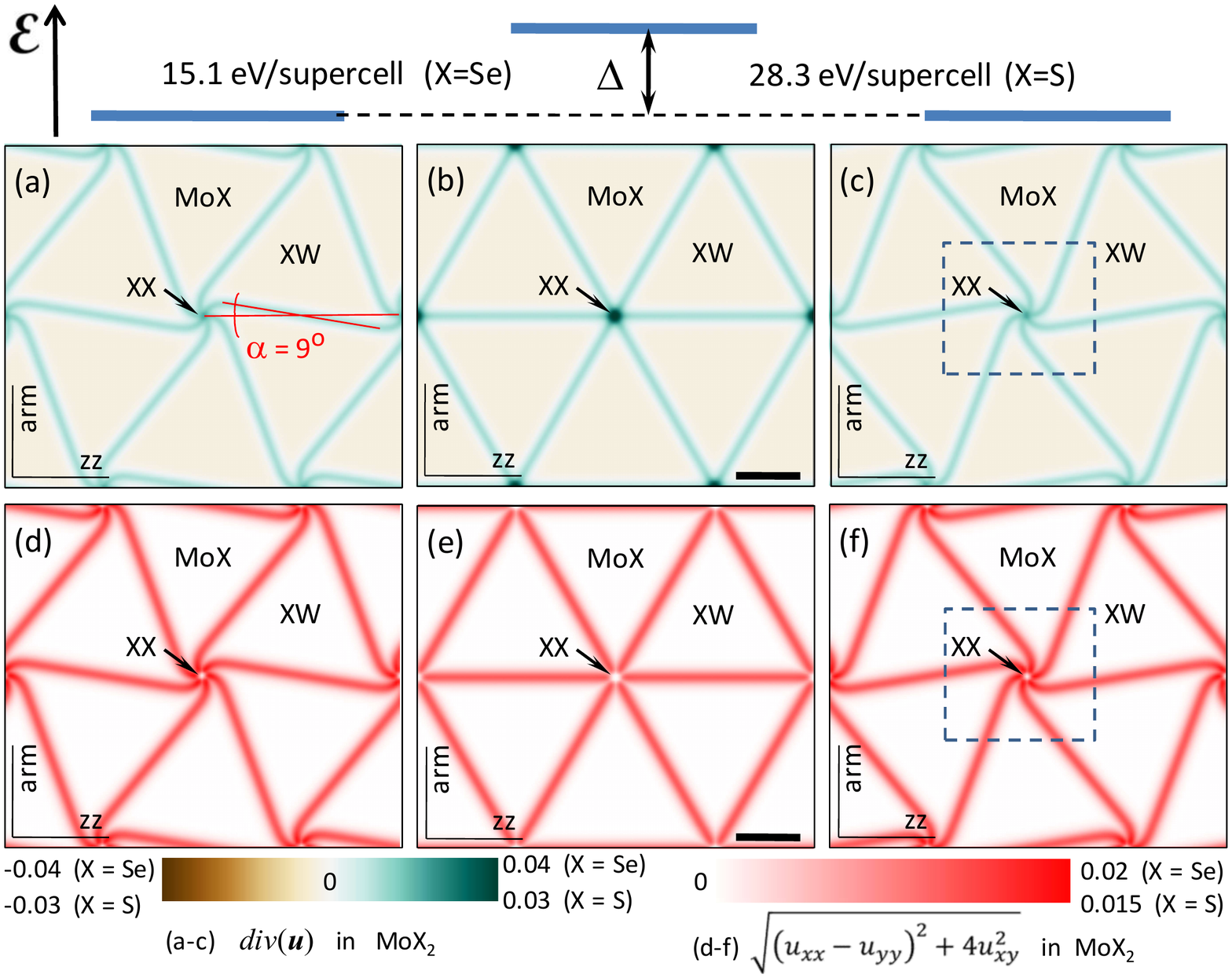}
\caption{\label{fig:2} Maps of hydrostatic (a-c) and shear (d-f) components of strain tensor in principal axes for symmetric (b) and twirled (a,c) structures of reconstructed moir\'e superlattice in aligned P-MoX$_2$/WX$_2$ bilayers. Sketch on the top shows energy gain from formation of twirled structures. Scale bar is 30\,nm and 45\,nm for X=Se and X=S, respectively. Dashed rectangles on (c) and (f) show area displayed in Fig. \ref{fig:3}.} 
\end{figure*}

\begin{figure}
\includegraphics[width=0.8\columnwidth]{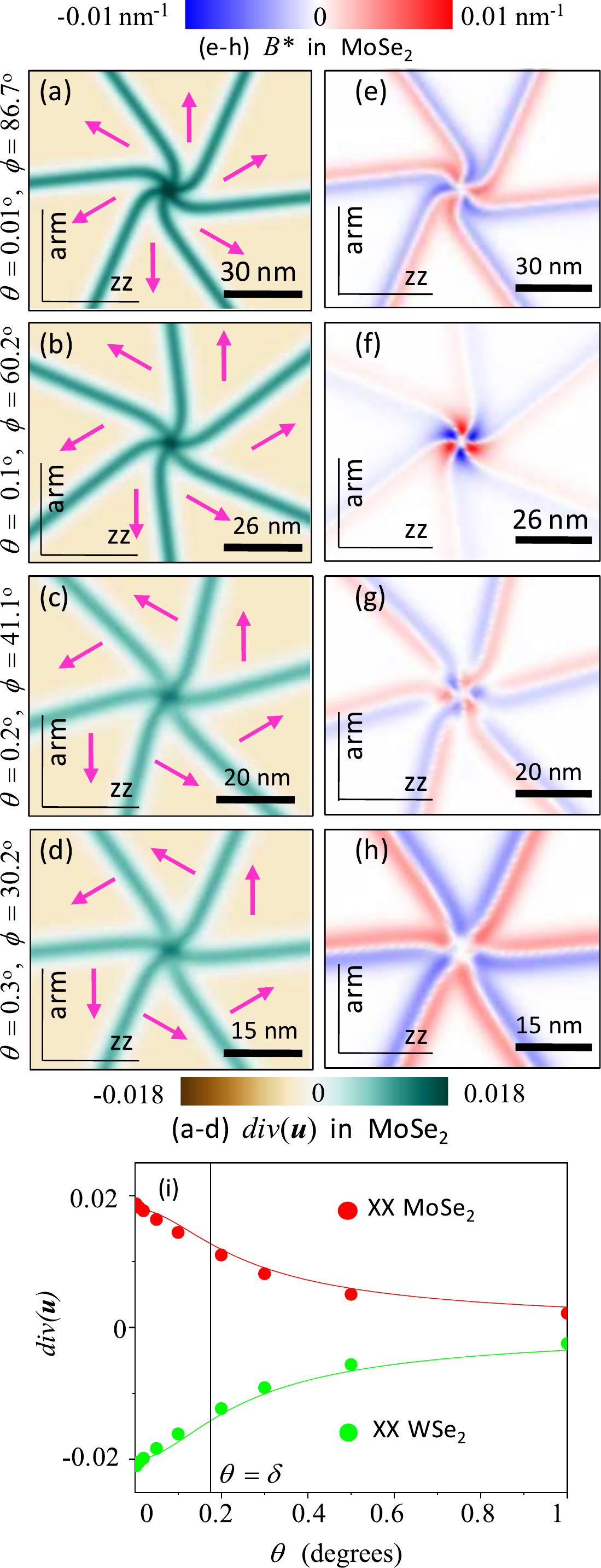}
\caption{\label{fig:3} Twirl in P-MoSe$_2$/WSe$_2$ for $\theta=0.01^{\circ}, 0.1^{\circ}, 0.2^{\circ}$ and $0.3^{\circ}$. (a-d) show maps of ${\rm div}\bm{u}_{\rm Mo}$. The arrows indicate direction of stacking vector $\bm{r}_0$ ($|\bm{r}_0|=a/\sqrt{3}$) inside domains; (e-h) show $B^* = 2\partial_x u^{\rm Mo}_{xy}+\partial_y(u^{\rm Mo}_{xx}-u^{\rm Mo}_{yy})$. (i) Values of ${\rm div}\bm{u}_{\rm Mo,W}$ at twirl center for various twist angles.
} 
\end{figure}

To find optimal distributions of displacement fields, $\bm{u}^{\rm W}$ and $\bm{u}^{\rm Mo}$, we solve a system of Lagrange-Euler equations, with periodic boundary conditions, implemented via the finite difference method. In particular, we chose rectangular supercell with $\ell\sqrt{3}/2$ and $\ell$ sides (\mbox{$\ell=a/\sqrt{\delta^2+\theta^2}$} is the moir\'e superlattice period, $a$ is averaged monolayer lattice parameter), setting the following boundary conditions: \mbox{$\bm{u}^{\rm Mo/W}(0,y')=\bm{u}^{\rm Mo/W}(\ell\sqrt{3}/2,y'\mp\ell/2)$}, at \mbox{$0\leq \pm y'\leq\ell/2$}, and
\mbox{$\bm{u}^{\rm Mo/W}(x',\ell/2)=\bm{u}^{\rm Mo/W}(x',-\ell/2)$}. Here, prime superscripts indicate that $x'Oy'$ reference frame was rotated by an angle $\phi = \pi/6 + {\rm arctan}\left[(\sqrt{3} \delta-\theta)/(\sqrt{3} \theta +\delta)\right]$ with respect to the fixed frame for which $Ox$ and $Oy$ axes are along zigzag and armchair directions in the crystal, respectively. Such a rotation allows us to fix boundary conditions (note that the Lagrange-Euler equations themselves are rotational-invariant). 

In the numerical analysis we use a sufficiently dense grid (one point per nm) to provide convergence of the computed displacement fields. This requires solving $\sim 30000$ coupled nonlinear equations. This solution is obtained in an annealing-type computational scheme: first, we choose the starting point as $\bm{u}^{\rm Mo/W}=\bm{0}$ and scale down the adhesion energy parameters $w^{(s,a)}_{1,2,3}$ by a small factor $\eta=10^{-3}$, finding slightly relaxed structure. Then, we gradually enlarge this factor, up to $\eta=1$ (at the final step), using solutions obtained at the previous iterations as starting points at each next step.

{\it Twirls in P-MoX$_2$/WX$_2$ heterostructures.} 
Lattice relaxation in P-MoX$_2$/WX$_2$ heterostructures results in formation of DWN separating triangular domains with MoX and XW stackings, similar to those found in 3R-TMD polytypes \cite{Weston2020,rosenberger2020} (see Fig. \ref{fig:2}). Minimising functional \eqref{Eq:min_functional} we find the following three competing distributions of strain. One of them (symmetric), shown on the middle panels, is characterized by straight edge dislocation lines linking XX stacking nodes of DWN. The other two (left and right panels) are twirled (broken-symmetry) structures of DWN nodes with a left/right-handed twist. The computed total energies per supercell, gathered in Table \ref{Tab:Penergies}, show that the twirled DWN structure has lower energy than the symmetric one. Note that  left/right-handed twirls have the same energy suggesting that DWN in perfectly aligned heterostructure ($\theta=0^{\circ}$) undergoes a spontaneous symmetry breaking into twirled state. 

Values of adhesive and elastic energies compared in Table \ref{Tab:Penergies} also suggest that the dominant energy gain for the twirled structures comes from lowering elastic energy. The latter is determined by hydrostatic strain, ${\rm div} \bm{u}$, and shear deformations, characterized by a vector, ${\bf A}=(u_{xx}-u_{yy},-2u_{xy})$. Their distribution across DWN is shown Fig. \ref{fig:2}, in the form of color maps for ${\rm div} \bm{u}$ and $|{\bf A}|$. Here, the difference between symmetric and twirled structures is such that hydrostatic strain component, concentrated around nodes, is higher for symmetric one.  Since energy costs of the hydrostatic strain  ($\propto\lambda+\mu$) are higher than those of shear strain ($\propto\mu$), the twirled structures emerge as energetically favourable.

{\setlength{\tabcolsep}{2.5pt}
\begin{table}[t]
\begin{center}
{\small	\caption{Values of adhesion, elastic, total energies (in eV/supercell) computed for symmetric (sym) and twirled (twirl) moir\'e superlattice structures and their total energy differences, $\Delta$, in aligned P-MoX$_2$/WX$_2$ bilayers. \label{Tab:Penergies}}}
	{\small
	\begin{tabular}{c|c|cccc}		
		\hline
		\hline
      X & structure & elastic & adhesion & total & $\Delta$ \\
        \hline
       \multirow{2}{*} {Se} & sym &  223.01  &  -2349.68 &  -2126.67 &\multirow{2}{*} {15.12} \\
                              & twirl & 208.95 &  -2350.74 &  -2141.79 & \\
        \hline
       \multirow{2}{*} {S} & sym &  349.92 &  -5321.11 & -4971.19  &\multirow{2}{*} {28.26} \\
                              & twirl & 323.33 &  -5322.78 & -4999.45 & \\
		\hline
		\hline
	\end{tabular}
	}
	\end{center}
\end{table}
}

In Fig. \ref{fig:3} we analyse dependence of ${\rm div} \bm{u}$ and ${\bf A}$ on the twist angle. The data shown in Fig. \ref{fig:3}(i) indicate that the maximal value of ${\rm div} \bm{u}$ at the hot spot of strain decreases with the twist, which also explains why twirling is weaker for larger $\theta$'s. Another deformation field characteristic which is interesting to consider is \mbox{$B^*=[{\rm rot}{\bf A}]_z=2\partial_x u^{\rm Mo(W)}_{xy}+\partial_y(u^{\rm Mo(W)}_{xx}-u^{\rm Mo(W)}_{yy})$}. This characteristic determines the size of piezoelectric charges, generated by inhomogeneous strain in each layer, and pseudomagnetic field that would be experienced by charge carriers at the K-valley band edge in the heterostructure \cite{Enaldiev_PRL,Enaldiev_2021}. Note that in P-heterostructures piezocharges ($\propto e_{11}B^*$) have opposite signs in MoX$_2$ and WX$_2$ layers, as $B^*_{\rm Mo}\approx-B^*_{\rm W}$, and piezocoefficients have the same signs, \mbox{$e_{11}^{\rm Mo}\approx e_{11}^{\rm W}$}. For the displayed range of angles, \mbox{$0^{\circ}\leq \theta\lesssim 0.4^{\circ}$}, distributions of $B^*$ shows change of sign and zero value in the middle of the domain wall. For larger twist angles $\theta\gg\delta$ piezocharge and pseudomagnetic field distributions take the form of those established earlier for homobilayers \cite{Enaldiev_PRL}. 

\begin{figure*}
\includegraphics[width=2.0\columnwidth]{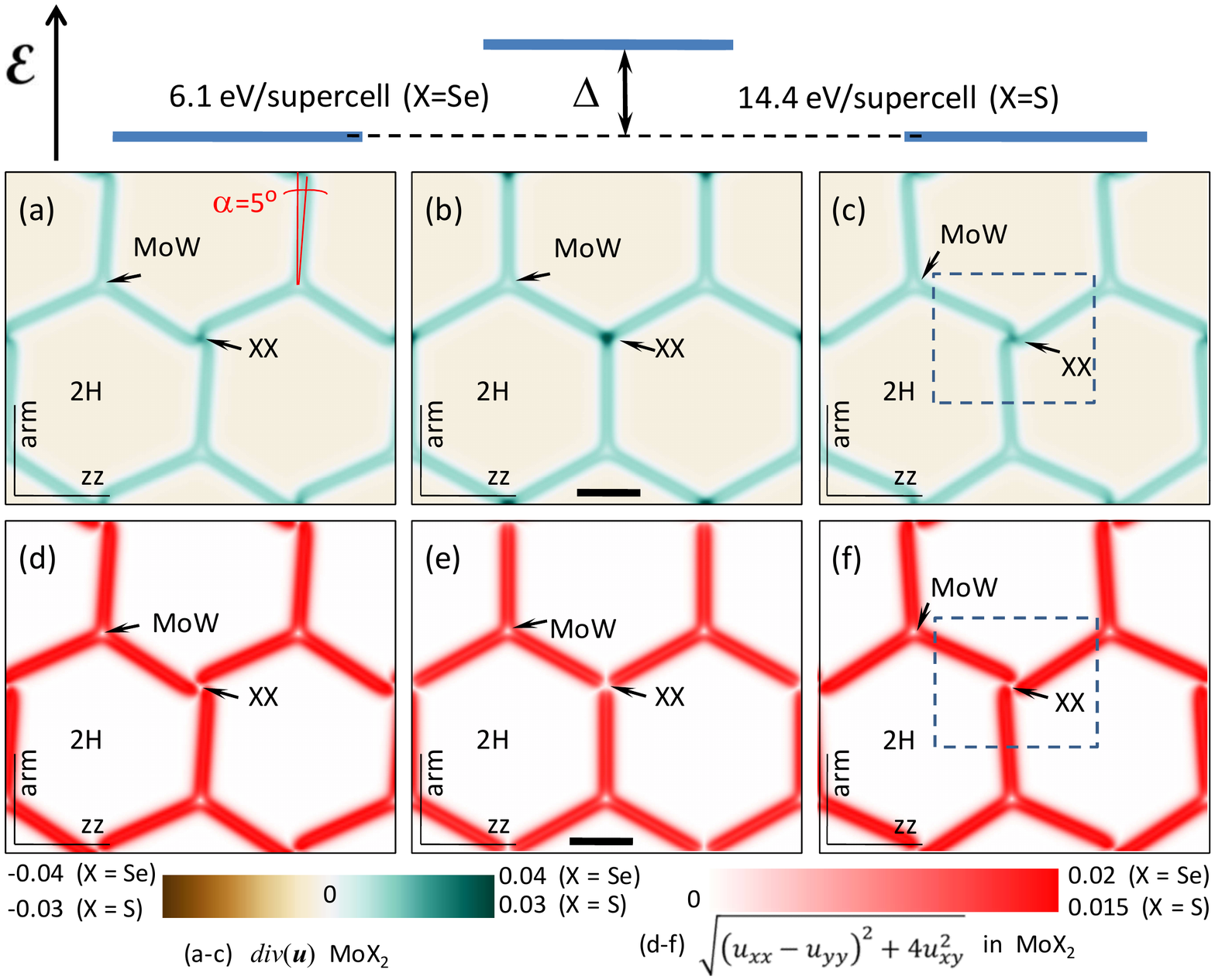}
\caption{\label{fig:4}  Maps of hydrostatic (a-c) and shear (d-f) components of strain tensor in principal axes for symmetric (b) and twirled (a,c) structures of reconstructed moir\'e superlattice in aligned AP-MoX$_2$/WX$_2$ bilayers. Sketch on the top shows energy gain from formation of twirled structures. Scale bar is 30\,nm and 45\,nm for X=Se and X=S, respectively. Dashed rectangles on (c) and (f) show area displayed in Fig. \ref{fig:5}.} 
\end{figure*}

{\it Twirls in AP-MoX$_2$/WX$_2$ heterostructures.}
Lattice relaxation in AP-MoX$_2$/WX$_2$ heterostructures leads to the formation of hexagonal domains of 2H-like stacking (simultaneous metal-on-chalcogen and chalcogen-on-metal like in bulk 2H TMD crystals). Those domains are separated by a hexagonal DWN with two inequivalent nodes: one with XX stacking (chalcogen-on-chalcogen) and the other with two metallic site on the top of each other, MoW. Upon minimising energy in Eq. \eqref{Eq:min_functional} for crystallographically aligned bilayers, we identify two candidates for the lowest energy configuration of DWN: symmetric and left/right-handed twirled structures, Fig. \ref{fig:4}. Twirled structures form around XX nodes, whereas DWN around MoW nodes is almost unchanged (only rotated as a whole by $\approx5^{\circ}$), which is because those nodes host metastable MoW stacking with much weaker hydrostatic strain component as compared to XX nodes). Energies of symmetric and twirled structures compared in Table \ref{Tab:APenergies} show that the broken-symmetry structure is energetically favourable. 

\begin{figure*}
\includegraphics[width=2.0\columnwidth]{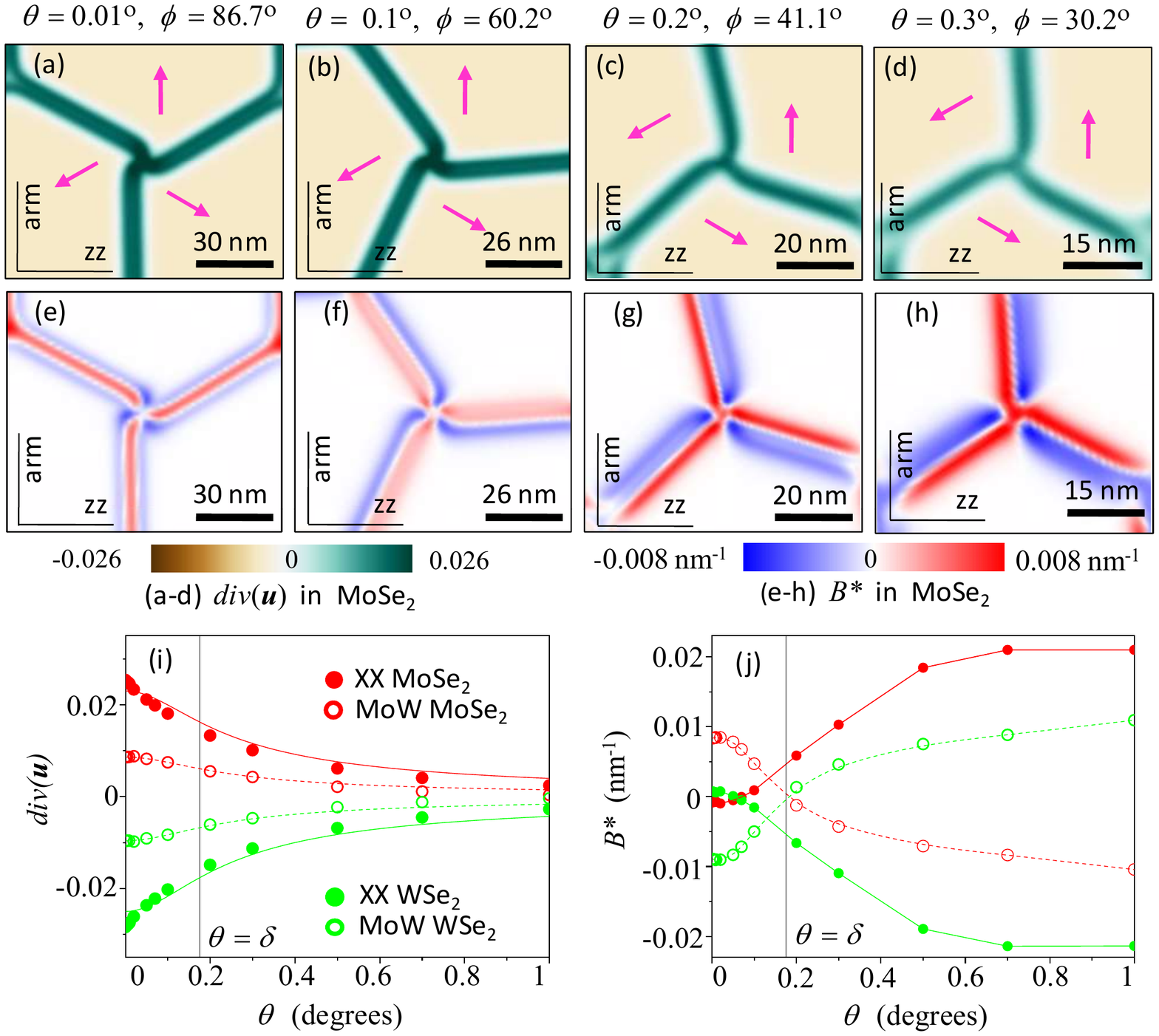}
\caption{\label{fig:5} 
Twirl in AP-MoSe$_2$/WSe$_2$ for $\theta=0.01^{\circ}, 0.1^{\circ}, 0.2^{\circ}$ and $0.3^{\circ}$. (a-d) show maps of ${\rm div}\bm{u}_{\rm Mo}$. The arrows indicate direction of stacking vector $\bm{r}_0$ ($|\bm{r}_0|=a/\sqrt{3}$) inside domains; (e-h) show $B^* = 2\partial_x u^{\rm Mo}_{xy}+\partial_y(u^{\rm Mo}_{xx}-u^{\rm Mo}_{yy})$. (i) Values of ${\rm div}\bm{u}_{\rm Mo,W}$ at twirl center for various twist angles. (j) $B^*$-dependence in centers of XX (twirl) and MoW nodes.} 
\end{figure*}

In Fig. \ref{fig:5} we show twist angle dependences of strain field characteristics around a single twirl. The data for ${\rm div}\bm{u}$ in Fig. \ref{fig:5}(i) indicate decrease of its magnitude (in each of the layers) with the twist angle. This trend corresponds to the decay of twirling with growth of layers' misorientation (Fig. \ref{fig:5}(a-d)) as already discussed about P-heterostructures in the previous section. We also point out that, in contrast to P-bilayers, for AP-heterostructures shear strain results in the same signs of piezocharge densities ($\propto e_{11}^{\rm Mo}B^*_{\rm Mo}\approx e_{11}^{\rm W}B^*_{\rm W}$) in both layers (opposite signs of piezocoefficients, \mbox{$e_{11}^{\rm Mo}\approx -e_{11}^{\rm W}$}, due to anti-alignment of layers, are compensated by signs inversion for \mbox{$B^*_{\rm Mo}\approx-B^*_{\rm W}$}). Note that DWN nodes host area of non-zero $B^*$, which reverses sign at $\theta\approx\delta$, with magnitude at MoW nodes substantially higher than that of twirled nodes (see Fig. \ref{fig:5}(j)).

{\setlength{\tabcolsep}{2.5pt}
\begin{table}[t]
\begin{center}
{\small	\caption{Values of adhesion, elastic, total energies (in eV/supercell) computed for symmetric (sym) and twirled (twirl) structures and total energy differences, $\Delta$, for aligned AP-MoX$_2$/WX$_2$ bilayers. \label{Tab:APenergies}}}
	{\small
	\begin{tabular}{c|c|cccc}		
		\hline
		\hline
        X& structure & elastic & adhesion & total & $\Delta$ \\
        \hline
        \multirow{2}{*} {Se} & sym &  329.46 &   -2648.55 &  -2319.09 &\multirow{2}{*} {6.12} \\
                              & twirl & 323.77 &  -2648.98 &  -2325.21 & \\
        \hline
        \multirow{2}{*} {S} & sym &  496.25 &   -5904.35 &  -5408.10 &\multirow{2}{*} {14.44}\\
         & twirl & 482.70 &  -5905.24 &  -5422.54 &   \\
		\hline
		\hline
	\end{tabular}
	}
	\end{center}
\end{table}
}

{\it Conclusion.}
The main finding of this study consists in the prediction of a spontaneous symmetry breaking in the form of domain wall network in MoX$_2$/WX$_2$ heterostructures. For almost perfectly aligned  MoX$_2$/WX$_2$ bilayers, this symmetry breaking consists in the formation of twirl-shaped nodes, accompanied by substantial reduction of a hydrostatic strain component in the vicinity of those nodes - as compared to earlier-studied symmetric (non-twirled) DWN structures \cite{Enaldiev2022a}. The reduction of hydrostatic strain component by twirling could have a pronounced effect in the energetics of the band edge states of the electrons and holes, making those less confined at the twirled nodes, as compared to symmetric nodes. As a result, the earlier studies \cite{Enaldiev2022a} overestimated electron, hole and exciton bindings by the XX DWN nodes, so that the analysis of electron, hole and interlayer excitons localisation at the twirls requires farther studies.

{\it Acknowledgement.} We acknowledge useful discussions with Marek Potemski, Isaac Ochoa and Fabio Ferreira. This work was supported by EC-FET Core 3 European Graphene Flagship Project, EC-FET Quantum Flagship Project 2D-SIPC, EPSRC grants EP/S030719/1 and EP/V007033/1, and the Lloyd Register Foundation Nanotechnology Grant.

{\it Author Contributions.} MAK, VVE, and VIF contributed equally to this paper. 
MAK and VVE developed formalism and software and carried out calculaions. VIF conceived the paper.
All authors contributed to the data analysis and manuscript writing.

\bibliography{bibl}

\begin{thebibliography}{12}%
\makeatletter
\providecommand \@ifxundefined [1]{%
 \@ifx{#1\undefined}
}%
\providecommand \@ifnum [1]{%
 \ifnum #1\expandafter \@firstoftwo
 \else \expandafter \@secondoftwo
 \fi
}%
\providecommand \@ifx [1]{%
 \ifx #1\expandafter \@firstoftwo
 \else \expandafter \@secondoftwo
 \fi
}%
\providecommand \natexlab [1]{#1}%
\providecommand \enquote  [1]{``#1''}%
\providecommand \bibnamefont  [1]{#1}%
\providecommand \bibfnamefont [1]{#1}%
\providecommand \citenamefont [1]{#1}%
\providecommand \href@noop [0]{\@secondoftwo}%
\providecommand \href [0]{\begingroup \@sanitize@url \@href}%
\providecommand \@href[1]{\@@startlink{#1}\@@href}%
\providecommand \@@href[1]{\endgroup#1\@@endlink}%
\providecommand \@sanitize@url [0]{\catcode `\\12\catcode `\$12\catcode
  `\&12\catcode `\#12\catcode `\^12\catcode `\_12\catcode `\%12\relax}%
\providecommand \@@startlink[1]{}%
\providecommand \@@endlink[0]{}%
\providecommand \url  [0]{\begingroup\@sanitize@url \@url }%
\providecommand \@url [1]{\endgroup\@href {#1}{\urlprefix }}%
\providecommand \urlprefix  [0]{URL }%
\providecommand \Eprint [0]{\href }%
\providecommand \doibase [0]{https://doi.org/}%
\providecommand \selectlanguage [0]{\@gobble}%
\providecommand \bibinfo  [0]{\@secondoftwo}%
\providecommand \bibfield  [0]{\@secondoftwo}%
\providecommand \translation [1]{[#1]}%
\providecommand \BibitemOpen [0]{}%
\providecommand \bibitemStop [0]{}%
\providecommand \bibitemNoStop [0]{.\EOS\space}%
\providecommand \EOS [0]{\spacefactor3000\relax}%
\providecommand \BibitemShut  [1]{\csname bibitem#1\endcsname}%
\let\auto@bib@innerbib\@empty
\bibitem [{\citenamefont {Enaldiev}\ \emph {et~al.}(2020)\citenamefont
  {Enaldiev}, \citenamefont {Z\'olyomi}, \citenamefont {Yelgel}, \citenamefont
  {Magorrian},\ and\ \citenamefont {Fal'ko}}]{Enaldiev_PRL}%
  \BibitemOpen
  \bibfield  {author} {\bibinfo {author} {\bibfnamefont {V.~V.}\ \bibnamefont
  {Enaldiev}}, \bibinfo {author} {\bibfnamefont {V.}~\bibnamefont {Z\'olyomi}},
  \bibinfo {author} {\bibfnamefont {C.}~\bibnamefont {Yelgel}}, \bibinfo
  {author} {\bibfnamefont {S.~J.}\ \bibnamefont {Magorrian}},\ and\ \bibinfo
  {author} {\bibfnamefont {V.~I.}\ \bibnamefont {Fal'ko}},\ }\bibfield  {title}
  {\bibinfo {title} {Stacking domains and dislocation networks in marginally
  twisted bilayers of transition metal dichalcogenides},\ }\href
  {https://doi.org/10.1103/PhysRevLett.124.206101} {\bibfield  {journal}
  {\bibinfo  {journal} {Phys. Rev. Lett.}\ }\textbf {\bibinfo {volume} {124}},\
  \bibinfo {pages} {206101} (\bibinfo {year} {2020})}\BibitemShut {NoStop}%
\bibitem [{\citenamefont {Weston}\ \emph {et~al.}(2020)\citenamefont {Weston},
  \citenamefont {Zou}, \citenamefont {Enaldiev}, \citenamefont {Summerfield},
  \citenamefont {Clark}, \citenamefont {Z{\'o}lyomi}, \citenamefont {Graham},
  \citenamefont {Yelgel}, \citenamefont {Magorrian}, \citenamefont {Zhou},
  \citenamefont {Zultak}, \citenamefont {Hopkinson}, \citenamefont {Barinov},
  \citenamefont {Bointon}, \citenamefont {Kretinin}, \citenamefont {Wilson},
  \citenamefont {Beton}, \citenamefont {Fal'ko}, \citenamefont {Haigh},\ and\
  \citenamefont {Gorbachev}}]{Weston2020}%
  \BibitemOpen
  \bibfield  {author} {\bibinfo {author} {\bibfnamefont {A.}~\bibnamefont
  {Weston}}, \bibinfo {author} {\bibfnamefont {Y.}~\bibnamefont {Zou}},
  \bibinfo {author} {\bibfnamefont {V.}~\bibnamefont {Enaldiev}}, \bibinfo
  {author} {\bibfnamefont {A.}~\bibnamefont {Summerfield}}, \bibinfo {author}
  {\bibfnamefont {N.}~\bibnamefont {Clark}}, \bibinfo {author} {\bibfnamefont
  {V.}~\bibnamefont {Z{\'o}lyomi}}, \bibinfo {author} {\bibfnamefont
  {A.}~\bibnamefont {Graham}}, \bibinfo {author} {\bibfnamefont
  {C.}~\bibnamefont {Yelgel}}, \bibinfo {author} {\bibfnamefont
  {S.}~\bibnamefont {Magorrian}}, \bibinfo {author} {\bibfnamefont
  {M.}~\bibnamefont {Zhou}}, \bibinfo {author} {\bibfnamefont {J.}~\bibnamefont
  {Zultak}}, \bibinfo {author} {\bibfnamefont {D.}~\bibnamefont {Hopkinson}},
  \bibinfo {author} {\bibfnamefont {A.}~\bibnamefont {Barinov}}, \bibinfo
  {author} {\bibfnamefont {T.~H.}\ \bibnamefont {Bointon}}, \bibinfo {author}
  {\bibfnamefont {A.}~\bibnamefont {Kretinin}}, \bibinfo {author}
  {\bibfnamefont {N.~R.}\ \bibnamefont {Wilson}}, \bibinfo {author}
  {\bibfnamefont {P.~H.}\ \bibnamefont {Beton}}, \bibinfo {author}
  {\bibfnamefont {V.~I.}\ \bibnamefont {Fal'ko}}, \bibinfo {author}
  {\bibfnamefont {S.~J.}\ \bibnamefont {Haigh}},\ and\ \bibinfo {author}
  {\bibfnamefont {R.}~\bibnamefont {Gorbachev}},\ }\bibfield  {title} {\bibinfo
  {title} {Atomic reconstruction in twisted bilayers of transition metal
  dichalcogenides},\ }\href {https://doi.org/10.1038/s41565-020-0682-9}
  {\bibfield  {journal} {\bibinfo  {journal} {Nature Nanotechnology}\ }\textbf
  {\bibinfo {volume} {15}},\ \bibinfo {pages} {592} (\bibinfo {year}
  {2020})}\BibitemShut {NoStop}%
\bibitem [{\citenamefont {Rosenberger}\ \emph {et~al.}(2020)\citenamefont
  {Rosenberger}, \citenamefont {Chuang}, \citenamefont {Phillips},
  \citenamefont {Oleshko}, \citenamefont {McCreary}, \citenamefont {Sivaram},
  \citenamefont {Hellberg},\ and\ \citenamefont {Jonker}}]{rosenberger2020}%
  \BibitemOpen
  \bibfield  {author} {\bibinfo {author} {\bibfnamefont {M.~R.}\ \bibnamefont
  {Rosenberger}}, \bibinfo {author} {\bibfnamefont {H.-J.}\ \bibnamefont
  {Chuang}}, \bibinfo {author} {\bibfnamefont {M.}~\bibnamefont {Phillips}},
  \bibinfo {author} {\bibfnamefont {V.~P.}\ \bibnamefont {Oleshko}}, \bibinfo
  {author} {\bibfnamefont {K.~M.}\ \bibnamefont {McCreary}}, \bibinfo {author}
  {\bibfnamefont {S.~V.}\ \bibnamefont {Sivaram}}, \bibinfo {author}
  {\bibfnamefont {C.~S.}\ \bibnamefont {Hellberg}},\ and\ \bibinfo {author}
  {\bibfnamefont {B.~T.}\ \bibnamefont {Jonker}},\ }\bibfield  {title}
  {\bibinfo {title} {Twist angle-dependent atomic reconstruction and moir{\'e}
  patterns in transition metal dichalcogenide heterostructures},\ }\href
  {https://doi.org/10.1021/acsnano.0c00088} {\bibfield  {journal} {\bibinfo
  {journal} {ACS Nano}\ }\textbf {\bibinfo {volume} {14}},\ \bibinfo {pages}
  {4550} (\bibinfo {year} {2020})}\BibitemShut {NoStop}%
\bibitem [{\citenamefont {Sung}\ \emph {et~al.}(2020)\citenamefont {Sung},
  \citenamefont {Zhou}, \citenamefont {Scuri}, \citenamefont {Z{\'o}lyomi},
  \citenamefont {Andersen}, \citenamefont {Yoo}, \citenamefont {Wild},
  \citenamefont {Joe}, \citenamefont {Gelly}, \citenamefont {Heo},
  \citenamefont {Magorrian}, \citenamefont {B{\'e}rub{\'e}}, \citenamefont
  {Valdivia}, \citenamefont {Taniguchi}, \citenamefont {Watanabe},
  \citenamefont {Lukin}, \citenamefont {Kim}, \citenamefont {Fal'ko},\ and\
  \citenamefont {Park}}]{Sung2020}%
  \BibitemOpen
  \bibfield  {author} {\bibinfo {author} {\bibfnamefont {J.}~\bibnamefont
  {Sung}}, \bibinfo {author} {\bibfnamefont {Y.}~\bibnamefont {Zhou}}, \bibinfo
  {author} {\bibfnamefont {G.}~\bibnamefont {Scuri}}, \bibinfo {author}
  {\bibfnamefont {V.}~\bibnamefont {Z{\'o}lyomi}}, \bibinfo {author}
  {\bibfnamefont {T.~I.}\ \bibnamefont {Andersen}}, \bibinfo {author}
  {\bibfnamefont {H.}~\bibnamefont {Yoo}}, \bibinfo {author} {\bibfnamefont
  {D.~S.}\ \bibnamefont {Wild}}, \bibinfo {author} {\bibfnamefont {A.~Y.}\
  \bibnamefont {Joe}}, \bibinfo {author} {\bibfnamefont {R.~J.}\ \bibnamefont
  {Gelly}}, \bibinfo {author} {\bibfnamefont {H.}~\bibnamefont {Heo}}, \bibinfo
  {author} {\bibfnamefont {S.~J.}\ \bibnamefont {Magorrian}}, \bibinfo {author}
  {\bibfnamefont {D.}~\bibnamefont {B{\'e}rub{\'e}}}, \bibinfo {author}
  {\bibfnamefont {A.~M.~M.}\ \bibnamefont {Valdivia}}, \bibinfo {author}
  {\bibfnamefont {T.}~\bibnamefont {Taniguchi}}, \bibinfo {author}
  {\bibfnamefont {K.}~\bibnamefont {Watanabe}}, \bibinfo {author}
  {\bibfnamefont {M.~D.}\ \bibnamefont {Lukin}}, \bibinfo {author}
  {\bibfnamefont {P.}~\bibnamefont {Kim}}, \bibinfo {author} {\bibfnamefont
  {V.~I.}\ \bibnamefont {Fal'ko}},\ and\ \bibinfo {author} {\bibfnamefont
  {H.}~\bibnamefont {Park}},\ }\bibfield  {title} {\bibinfo {title} {Broken
  mirror symmetry in excitonic response of reconstructed domains in twisted
  {MoSe}$_2$/{MoSe}$_2$ bilayers},\ }\href
  {https://doi.org/10.1038/s41565-020-0728-z} {\bibfield  {journal} {\bibinfo
  {journal} {Nature Nanotechnology}\ }\textbf {\bibinfo {volume} {15}},\
  \bibinfo {pages} {750} (\bibinfo {year} {2020})}\BibitemShut {NoStop}%
\bibitem [{\citenamefont {McGilly}\ \emph {et~al.}(2020)\citenamefont
  {McGilly}, \citenamefont {Kerelsky}, \citenamefont {Finney}, \citenamefont
  {Shapovalov}, \citenamefont {Shih}, \citenamefont {Ghiotto}, \citenamefont
  {Zeng}, \citenamefont {Moore}, \citenamefont {Wu}, \citenamefont {Bai},
  \citenamefont {Watanabe}, \citenamefont {Taniguchi}, \citenamefont {Stengel},
  \citenamefont {Zhou}, \citenamefont {Hone}, \citenamefont {Zhu},
  \citenamefont {Basov}, \citenamefont {Dean}, \citenamefont {Dreyer},\ and\
  \citenamefont {Pasupathy}}]{McGilly2020}%
  \BibitemOpen
  \bibfield  {author} {\bibinfo {author} {\bibfnamefont {L.~J.}\ \bibnamefont
  {McGilly}}, \bibinfo {author} {\bibfnamefont {A.}~\bibnamefont {Kerelsky}},
  \bibinfo {author} {\bibfnamefont {N.~R.}\ \bibnamefont {Finney}}, \bibinfo
  {author} {\bibfnamefont {K.}~\bibnamefont {Shapovalov}}, \bibinfo {author}
  {\bibfnamefont {E.-M.}\ \bibnamefont {Shih}}, \bibinfo {author}
  {\bibfnamefont {A.}~\bibnamefont {Ghiotto}}, \bibinfo {author} {\bibfnamefont
  {Y.}~\bibnamefont {Zeng}}, \bibinfo {author} {\bibfnamefont {S.~L.}\
  \bibnamefont {Moore}}, \bibinfo {author} {\bibfnamefont {W.}~\bibnamefont
  {Wu}}, \bibinfo {author} {\bibfnamefont {Y.}~\bibnamefont {Bai}}, \bibinfo
  {author} {\bibfnamefont {K.}~\bibnamefont {Watanabe}}, \bibinfo {author}
  {\bibfnamefont {T.}~\bibnamefont {Taniguchi}}, \bibinfo {author}
  {\bibfnamefont {M.}~\bibnamefont {Stengel}}, \bibinfo {author} {\bibfnamefont
  {L.}~\bibnamefont {Zhou}}, \bibinfo {author} {\bibfnamefont {J.}~\bibnamefont
  {Hone}}, \bibinfo {author} {\bibfnamefont {X.}~\bibnamefont {Zhu}}, \bibinfo
  {author} {\bibfnamefont {D.~N.}\ \bibnamefont {Basov}}, \bibinfo {author}
  {\bibfnamefont {C.}~\bibnamefont {Dean}}, \bibinfo {author} {\bibfnamefont
  {C.~E.}\ \bibnamefont {Dreyer}},\ and\ \bibinfo {author} {\bibfnamefont
  {A.~N.}\ \bibnamefont {Pasupathy}},\ }\bibfield  {title} {\bibinfo {title}
  {Visualization of moir\'e superlattices},\ }\href
  {https://doi.org/10.1038/s41565-020-0708-3} {\bibfield  {journal} {\bibinfo
  {journal} {Nature Nanotechnology}\ }\textbf {\bibinfo {volume} {15}},\
  \bibinfo {pages} {580} (\bibinfo {year} {2020})}\BibitemShut {NoStop}%
\bibitem [{\citenamefont {Shabani}\ \emph {et~al.}(2021)\citenamefont
  {Shabani}, \citenamefont {Halbertal}, \citenamefont {Wu}, \citenamefont
  {Chen}, \citenamefont {Liu}, \citenamefont {Hone}, \citenamefont {Yao},
  \citenamefont {Basov}, \citenamefont {Zhu},\ and\ \citenamefont
  {Pasupathy}}]{Shabani2021}%
  \BibitemOpen
  \bibfield  {author} {\bibinfo {author} {\bibfnamefont {S.}~\bibnamefont
  {Shabani}}, \bibinfo {author} {\bibfnamefont {D.}~\bibnamefont {Halbertal}},
  \bibinfo {author} {\bibfnamefont {W.}~\bibnamefont {Wu}}, \bibinfo {author}
  {\bibfnamefont {M.}~\bibnamefont {Chen}}, \bibinfo {author} {\bibfnamefont
  {S.}~\bibnamefont {Liu}}, \bibinfo {author} {\bibfnamefont {J.}~\bibnamefont
  {Hone}}, \bibinfo {author} {\bibfnamefont {W.}~\bibnamefont {Yao}}, \bibinfo
  {author} {\bibfnamefont {D.~N.}\ \bibnamefont {Basov}}, \bibinfo {author}
  {\bibfnamefont {X.}~\bibnamefont {Zhu}},\ and\ \bibinfo {author}
  {\bibfnamefont {A.~N.}\ \bibnamefont {Pasupathy}},\ }\bibfield  {title}
  {\bibinfo {title} {Deep moir{\'{e}} potentials in twisted transition metal
  dichalcogenide bilayers},\ }\href
  {https://doi.org/10.1038/s41567-021-01174-7} {\bibfield  {journal} {\bibinfo
  {journal} {Nature Physics}\ }\textbf {\bibinfo {volume} {17}},\ \bibinfo
  {pages} {720} (\bibinfo {year} {2021})}\BibitemShut {NoStop}%
\bibitem [{\citenamefont {Naik}\ and\ \citenamefont
  {Jain}(2018)}]{NaikPRL2018}%
  \BibitemOpen
  \bibfield  {author} {\bibinfo {author} {\bibfnamefont {M.~H.}\ \bibnamefont
  {Naik}}\ and\ \bibinfo {author} {\bibfnamefont {M.}~\bibnamefont {Jain}},\
  }\bibfield  {title} {\bibinfo {title} {Ultraflatbands and shear solitons in
  moir\'e patterns of twisted bilayer transition metal dichalcogenides},\
  }\href {https://doi.org/10.1103/PhysRevLett.121.266401} {\bibfield  {journal}
  {\bibinfo  {journal} {Phys. Rev. Lett.}\ }\textbf {\bibinfo {volume} {121}},\
  \bibinfo {pages} {266401} (\bibinfo {year} {2018})}\BibitemShut {NoStop}%
\bibitem [{\citenamefont {Carr}\ \emph {et~al.}(2018)\citenamefont {Carr},
  \citenamefont {Massatt}, \citenamefont {Torrisi}, \citenamefont {Cazeaux},
  \citenamefont {Luskin},\ and\ \citenamefont {Kaxiras}}]{CarrPRB2018}%
  \BibitemOpen
  \bibfield  {author} {\bibinfo {author} {\bibfnamefont {S.}~\bibnamefont
  {Carr}}, \bibinfo {author} {\bibfnamefont {D.}~\bibnamefont {Massatt}},
  \bibinfo {author} {\bibfnamefont {S.~B.}\ \bibnamefont {Torrisi}}, \bibinfo
  {author} {\bibfnamefont {P.}~\bibnamefont {Cazeaux}}, \bibinfo {author}
  {\bibfnamefont {M.}~\bibnamefont {Luskin}},\ and\ \bibinfo {author}
  {\bibfnamefont {E.}~\bibnamefont {Kaxiras}},\ }\bibfield  {title} {\bibinfo
  {title} {Relaxation and domain formation in incommensurate two-dimensional
  heterostructures},\ }\href {https://doi.org/10.1103/PhysRevB.98.224102}
  {\bibfield  {journal} {\bibinfo  {journal} {Phys. Rev. B}\ }\textbf {\bibinfo
  {volume} {98}},\ \bibinfo {pages} {224102} (\bibinfo {year}
  {2018})}\BibitemShut {NoStop}%
\bibitem [{\citenamefont {Enaldiev}\ \emph {et~al.}(2022)\citenamefont
  {Enaldiev}, \citenamefont {Ferreira}, \citenamefont {McHugh},\ and\
  \citenamefont {Fal'ko}}]{Enaldiev2022a}%
  \BibitemOpen
  \bibfield  {author} {\bibinfo {author} {\bibfnamefont {V.~V.}\ \bibnamefont
  {Enaldiev}}, \bibinfo {author} {\bibfnamefont {F.}~\bibnamefont {Ferreira}},
  \bibinfo {author} {\bibfnamefont {J.~G.}\ \bibnamefont {McHugh}},\ and\
  \bibinfo {author} {\bibfnamefont {V.~I.}\ \bibnamefont {Fal'ko}},\ }\bibfield
   {title} {\bibinfo {title} {Self-organized quantum dots in marginally twisted
  {MoSe}$_2$/{WSe}$_2$ and {MoS}$_2$/{WS}$_2$ bilayers},\ }\href
  {https://doi.org/10.1038/s41699-022-00346-0} {\bibfield  {journal} {\bibinfo
  {journal} {npj 2D Materials and Applications}\ }\textbf {\bibinfo {volume}
  {6}},\ \bibinfo {pages} {74} (\bibinfo {year} {2022})}\BibitemShut {NoStop}%
\bibitem [{\citenamefont {Enaldiev}\ \emph {et~al.}(2021)\citenamefont
  {Enaldiev}, \citenamefont {Ferreira}, \citenamefont {Magorrian},\ and\
  \citenamefont {Fal'ko}}]{Enaldiev_2021}%
  \BibitemOpen
  \bibfield  {author} {\bibinfo {author} {\bibfnamefont {V.~V.}\ \bibnamefont
  {Enaldiev}}, \bibinfo {author} {\bibfnamefont {F.}~\bibnamefont {Ferreira}},
  \bibinfo {author} {\bibfnamefont {S.~J.}\ \bibnamefont {Magorrian}},\ and\
  \bibinfo {author} {\bibfnamefont {V.~I.}\ \bibnamefont {Fal'ko}},\ }\bibfield
   {title} {\bibinfo {title} {Piezoelectric networks and ferroelectric domains
  in twistronic superlattices in {WS}$_2$/{MoS}$_2$ and {WSe}$_2$/{MoSe}$_2$
  bilayers},\ }\href {https://doi.org/10.1088/2053-1583/abdd92} {\bibfield
  {journal} {\bibinfo  {journal} {2D Materials}\ }\textbf {\bibinfo {volume}
  {8}},\ \bibinfo {pages} {025030} (\bibinfo {year} {2021})}\BibitemShut
  {NoStop}%
\bibitem [{\citenamefont {Igui{\~n}iz}\ \emph {et~al.}(2019)\citenamefont
  {Igui{\~n}iz}, \citenamefont {Frisenda}, \citenamefont {Bratschitsch},\ and\
  \citenamefont {Castellanos-Gomez}}]{iguiniz2019}%
  \BibitemOpen
  \bibfield  {author} {\bibinfo {author} {\bibfnamefont {N.}~\bibnamefont
  {Igui{\~n}iz}}, \bibinfo {author} {\bibfnamefont {R.}~\bibnamefont
  {Frisenda}}, \bibinfo {author} {\bibfnamefont {R.}~\bibnamefont
  {Bratschitsch}},\ and\ \bibinfo {author} {\bibfnamefont {A.}~\bibnamefont
  {Castellanos-Gomez}},\ }\bibfield  {title} {\bibinfo {title} {Revisiting the
  buckling metrology method to determine the young's modulus of 2d materials},\
  }\href@noop {} {\bibfield  {journal} {\bibinfo  {journal} {Advanced
  Materials}\ }\textbf {\bibinfo {volume} {31}},\ \bibinfo {pages} {1807150}
  (\bibinfo {year} {2019})}\BibitemShut {NoStop}%
\bibitem [{\citenamefont {Androulidakis}\ \emph {et~al.}(2018)\citenamefont
  {Androulidakis}, \citenamefont {Zhang}, \citenamefont {Robertson},\ and\
  \citenamefont {Tawfick}}]{androulidakis2018}%
  \BibitemOpen
  \bibfield  {author} {\bibinfo {author} {\bibfnamefont {C.}~\bibnamefont
  {Androulidakis}}, \bibinfo {author} {\bibfnamefont {K.}~\bibnamefont
  {Zhang}}, \bibinfo {author} {\bibfnamefont {M.}~\bibnamefont {Robertson}},\
  and\ \bibinfo {author} {\bibfnamefont {S.}~\bibnamefont {Tawfick}},\
  }\bibfield  {title} {\bibinfo {title} {Tailoring the mechanical properties of
  2d materials and heterostructures},\ }\href@noop {} {\bibfield  {journal}
  {\bibinfo  {journal} {2D Materials}\ }\textbf {\bibinfo {volume} {5}},\
  \bibinfo {pages} {032005} (\bibinfo {year} {2018})}\BibitemShut {NoStop}%
\end{thebibliography}%
\end{document}